\documentclass[prb,showpacs,superscriptaddress,amsmath,amssymb,floatfix,twocolumn]{revtex4}
\usepackage{graphicx}
\usepackage{dcolumn}
\usepackage{bm}
\usepackage{color}

\newcommand{\lsco}{La$_{2-x}$Sr$_x$CoO$_4$\,}
\newcommand{\lco}{La$_2$CoO$_4$\,}
\newcommand{\cppp}{Co$^{3+}$\,}
\newcommand{\cpp}{Co$^{2+}$\,}

\newcommand{\mbf}[1]{\mathbf #1}

\begin{document}

\title{Numerical Modelling of Disordered Magnetic Systems: {\lsco}}
\author{M.~Rotter} 
  \affiliation{martin\_rotter@mcphase.de, Neutron Science Laboratory, Institute for Solid State Physics , University of Tokyo, Tokai 319-1106, Japan /
McPhase Project www.mcphase.de, Dresden, Germany }

\date{\today}
\begin{abstract}
Magnetic properties of \lsco have been modelled using McPhase for different doping levels x.
Footprints of different interactions, such as local single ion anisotropy and variations in the
exchange interactions have been identified in the profile of the diffuse elastic neutron scattering cross section.
The role of quantum fluctuations in the formation of the hourglass spectrum is discussed.
\end{abstract}

\maketitle

\section{Introduction}

Charge, spin and orbital order in layered transition metal oxides  may be influenced considerably by
electron doping resulting in  exotic and complex physical properties, which have high potential for technological application.

This study focuses on \lsco , which  has been  studied experimentally recently. There is large consensus, that in this system
the charge order is static and stable up to at least room temperature~\cite{Zaliznyak00-4353,helme09-134414}. However, it is
matter of ongoing debate, if charge stripes are present in this system or not. For example, the famous hourglass spectrum for $x \sim 0.4$ can be 
interpreted based on a  (i) frustrated charge stripe~\cite{gaw13-165121,boothroyd11-341,andrade12-147201} and 
a  (ii) frustrated checkerboard charge order~\cite{drees13-3449,drees14-6731,guo15-580}.
Indeed, no clear experimental evidence for the existence of charge stripes in \lsco has been found and thus it is interesting to investigate by
model calculations, if the existence of charge stripes is indeed needed in order to interpret all available experimental data. 
In an attempt to answer this question, extensive simulations using 
the McPhase 5.2 software suite~\cite[www.mcphase.de]{rotter12-213201} have been performed. In the next section
the basic model will be  described, followed by a  presentation of selected calculation results.
The last section is devoted to the modelling of quantum fluctuations using a magnetic cluster
approach. The input files  for McPhase and summary logbooks of the calculations are available as a supplementary material to this article under
{\em www.mcphase.de/lsco.zip}

\section{Description of the Model}\label{model}

\subsection{Generation of frustrated Checkerboard Charge Order Pattern}\label{comodel}

Statistical/random disordered checkerboard charge order pattern of size 30x30 were generated from Monte-Carlo simulations similar to~\cite[Drees et al. and Andrade et al.]{drees14-6731,andrade12-147201}
using the Metropolis algorithm for the two dimensional Ising model. Only one potential for Coulomb repulsion of two adjacent Co$^{3+}$ and another for steric repulsion of  Co$^{2+}$ ions with large ionic radii has been considered.  A longitudinal external field was used to tune the number of  Co$^{3+}$  ions according to the 
desired doping level $x$. For the simulation of diffuse scattering an average of 64 configurations was performed.

\subsection{Modelling the Magnetic Properties}\label{magprop}

Magnetism was modeled by introducing antiferromagnetic exchange interactions between the Co$^{2+}$ ions $i,j$ with spin operators $\hat {\mbf  S}^i$ and  $\hat {\mbf S}^j$, respectively (the $Co^{3+}$ ions were assumed to be in the low spin state and treated as nonmagnetic):

\begin{equation}\label{Hex}
H_{ex}=- \frac{1}{2} \sum_{i,j} J(ij)  \hat {\mbf S}^i \cdot \hat {\mbf S}^j
\end{equation}

The simplest model  is obtained by introducing two exchange constants. We set $J(100)=-9.5$~meV 
as observed in the parent phase~\cite{babkevich10-184425} with $x=0$. Note, the ''100'' in $J(100)$ refers
to the pseudotetragonal unit cell of the high temperature tetragonal phase. In addition, 
$J(200)=-1.4$~meV is considered, if the interstitial (100) neighbour ion is \cppp, this
value is taken from the pure checkerboard charge ordered phase~\cite{helme09-134414} with $x=0.5$.

An effort was made to extend previous work on the dilute systems, where the \cpp spin had been modelled using an effective S=1/2. 
Here, the properties of the \cpp ion were calculated based on the same coupling scheme as used in~\cite{helme09-134414,babkevich10-184425} including
crystal field and spin orbit interaction parameters $B_2^0=14.6$~meV, $B_4^0=-1.35$~meV, $B_4^4=-8.0$~meV, $\lambda=-22.1$~meV , for notation see~\cite{hutchings64-227}. 


\begin{equation}\label{Hsi}
H_{Co}^i=\sum_{lm} B_l^m O_l^m(\hat {\mbf L}^i) + \lambda \hat {\mbf L}^i \cdot \hat {\mbf S}^i 
\end{equation}

Note that the crystal field anisotropy will tend to align the magnetic moments perpendicular to $c$.
In order to enable calculations involving many ions, the spin matrices of the
exchange Hamiltonian~(\ref{Hex}) were projected to the 4 lowest lying eigenstates $|\alpha \rangle$ ($\alpha =1,\dots,4$ with energies $E_{\alpha}$)
of the single ion Hamiltonian~(\ref{Hsi}) of the Co ions:

\begin{equation}\label{subs}
\hat {\mbf S}^i \sim \sum_{\alpha,\beta}|\alpha\rangle \langle\alpha|\hat {\mbf S}^i|\beta\rangle \langle\beta|
\end{equation}

With this substitution~(\ref{subs}) inserted in the exchange interaction~(\ref{Hex}) the total Hamiltonian becomes:

\begin{equation}\label{Ham}
H=H_{ex} + \sum_{i,\alpha} |\alpha\rangle E_{\alpha} \langle\alpha|
\end{equation}

\begin{figure}[htbp] 
  \centering
     \includegraphics[width=0.41\textwidth]{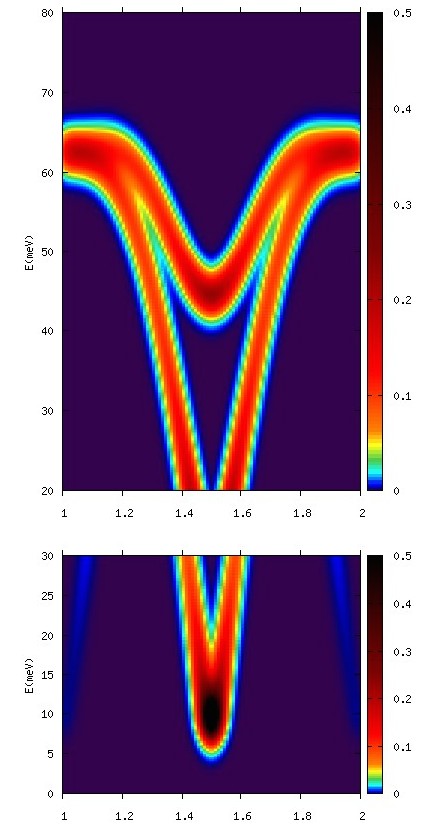}
     \includegraphics[width=0.41\textwidth]{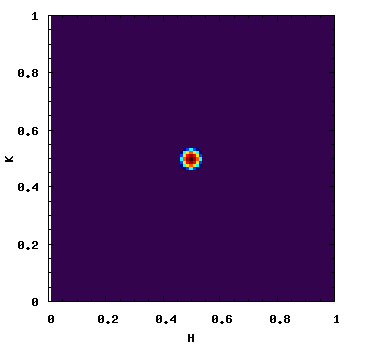}
  \caption{Simulated inelastic (top: along (h 0.5 0), middle: along (hh0)  and elastic (bottom: (hk0) plane)  magnetic neutron pattern for x=0. }
  \label{qq-x0p00}
\end{figure}
\begin{figure}[htbp] 
  \centering
     \includegraphics[width=0.41\textwidth]{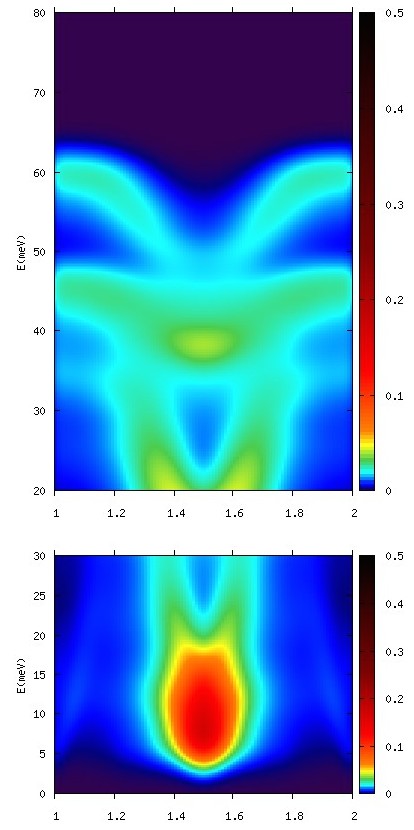}
     \includegraphics[width=0.41\textwidth]{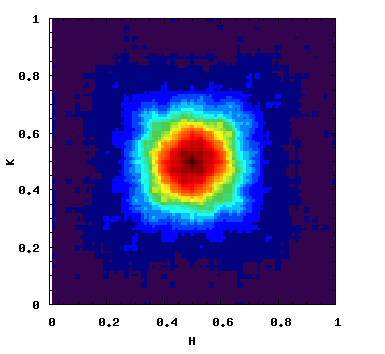}
  \caption{Simulated inelastic (top: along (h 0.5 0), middle: along (hh0)  and elastic (bottom: (hk0) plane)  magnetic neutron pattern for x=0.25 }
  \label{qq-x0p25}
\end{figure}
\begin{figure}[htbp] 
  \centering
     \includegraphics[width=0.41\textwidth]{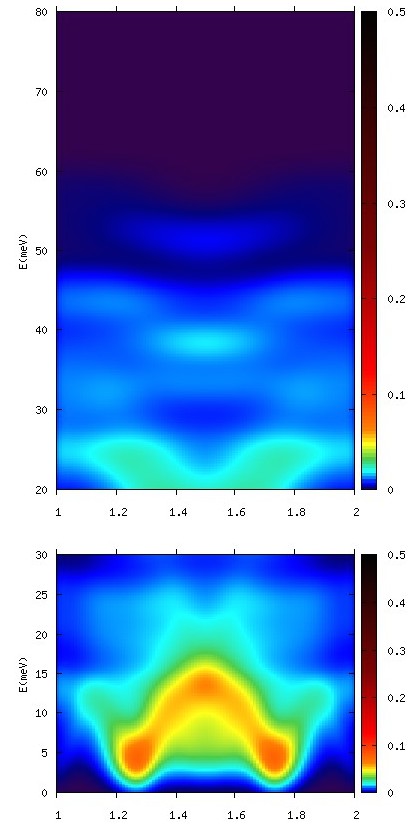}
     \includegraphics[width=0.41\textwidth]{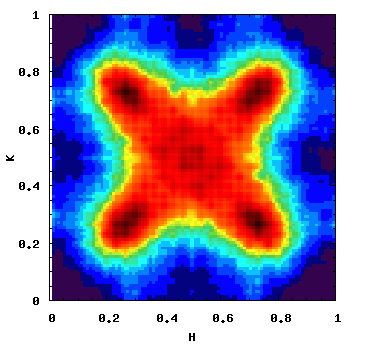}
  \caption{Simulated inelastic (top: along (h 0.5 0), middle: along (hh0)  and elastic (bottom: (hk0) plane)  magnetic neutron pattern for x=0.4 }
  \label{qq-x0p40}
\end{figure}
\begin{figure}[htbp] 
  \centering
     \includegraphics[width=0.41\textwidth]{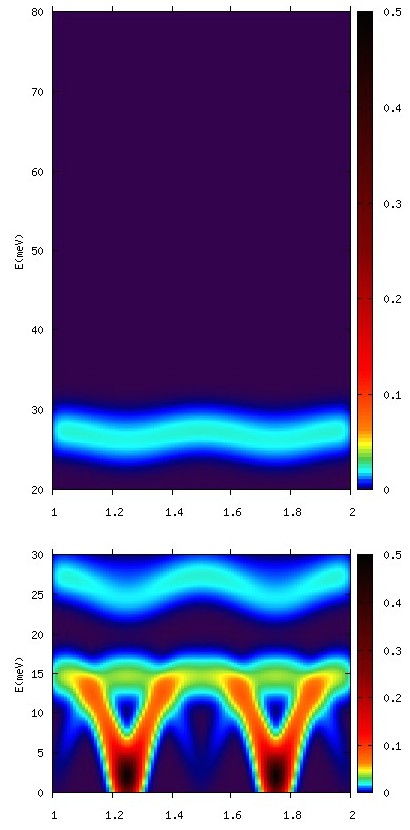}
     \includegraphics[width=0.41\textwidth]{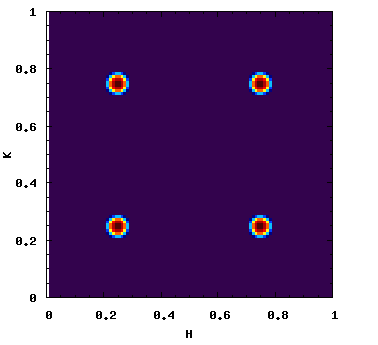}
  \caption{Simulated inelastic (top: along (h 0.5 0), middle: along (hh0)  and elastic (bottom: (hk0) plane)  magnetic neutron pattern for x=0.5 }
  \label{qq-x0p50}
\end{figure}
\begin{figure}[htbp] 
  \centering
     \includegraphics[width=0.41\textwidth]{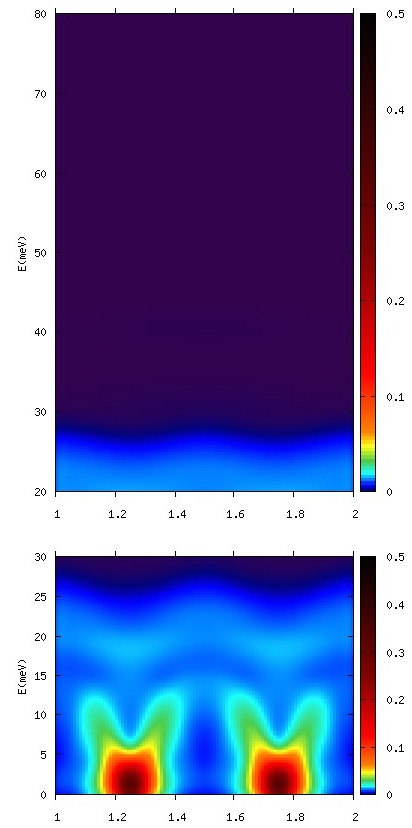}
     \includegraphics[width=0.41\textwidth]{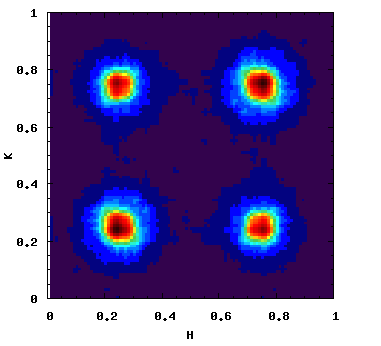}
  \caption{Simulated inelastic (top: along (h 0.5 0), middle: along (hh0)  and elastic (bottom: (hk0) plane)  magnetic neutron pattern for x=0.6 }
  \label{qq-x0p60}
\end{figure}

\begin{figure}[htbp] 
  \centering
     \includegraphics[width=0.5\textwidth]{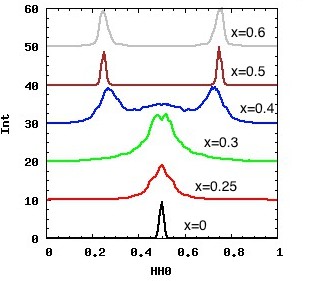}
  \caption{Simulated elastic magnetic neutron pattern along hh0 for different $x$. }
  \label{qq-hh0}
\end{figure}

In several figures of this article calculated neutron scattering cross sections are shown in dipole approximation.
Energies and Intensities of the modes were evaluated by dynamical matrix diagonalisation (for 
details see Rotter et al.~\cite{rotter12-213201}).
The general expression for the double differential magnetic scattering cross section 
for unpolarised neutrons has been given frequently in literature (see e.g. \cite{lovesey84-1}):

\begin{eqnarray}\label{dsdoderepeated}
\frac{d^2\sigma_{\rm mag}}{d\Omega dE'}=\frac{k'}{k}\left( \frac{ \gamma r_0}{2 \mu_B}  \right)^2
\sum_{\alpha\beta=1,2,3}(\delta_{\alpha\beta}- \frac{Q_{\alpha} Q_{\beta}}{|{\mbf Q}|^2})\times \nonumber\\
\frac{1}{2\pi\hbar}\int_{-\infty}^{+\infty}dt e^{i\omega t}
\sum_{nn'} e^{-W_n(Q)- W_{n'}(Q)} e^{-i\mbf Q \cdot (\mbf R_n-\mbf R_{n'})} \nonumber \\
\times \langle \hat M_{\alpha}^{n \dag}(t,\mbf Q)  \hat M_{\beta}^{n'}(0,\mbf Q) \rangle_{T,H}
\end{eqnarray}

In (\ref{dsdoderepeated}) $\mbf k$ and $\mbf k'$ denote the wave vector 
of the incoming and scattered neutron, respectively.
The total magnetic cross section is $4\pi (\gamma r_0)^2=4\pi\left(\frac{\hbar \gamma e^2}{mc^2}\right)^2
=3.65$~barn. 
$\hbar \omega=E-E'$  and
$\mbf Q =\mbf k-\mbf k'$  denote the energy and momentum transfer.
 $\exp(-W_n(Q))$ is  the Debye-Waller
factor of the atom number $n$.

The magnetic scattering operator $ \hat {\mbf M}({\mbf Q})$ was obtained from the spin $\hat {\mbf S}$ by applying an anisotropic
Land\'e factor $g_x=g_y=2.79$, $g_z=2$.
 In the calculation of
the cross section the form factor and the Debye Waller factors were not considered (set to 1), i.e.

\begin{equation}
 \hat  M_{\alpha}({\mbf Q}) \sim  g_{\alpha} \hat S_{\alpha}
\end{equation}

The resolution function for elastic scattering was assumed to be Gaussian with a width of $\Delta h = \Delta k=0.03$ in both h and k direction,
for inelastic scattering a Q-resolution of $\Delta h=0.05$ and an energy resolution $\Delta E=4$~meV was used.

In this way it is not only possible to model the hourglass spectrum for $x=0.4$, but also to
investigate it's evolvement for different $x$ quantitatively. For example, the 
optical modes observed for $x=0.5$ may be modeled (not possible within an effective S=1/2 approach).
Figs.~\ref{qq-x0p00} to ~\ref{qq-hh0} show the results of such calculations.

Still, there are major drawbacks of this model:
\begin{itemize}
\item Note that the gap in the
excitation spectra was produced by introducing an anisotropy field on each \cpp site similar to~\cite{helme09-134414,babkevich10-184425}: the calculated
mean fields were simply increased by 1\%. However, the source of this anisotropy field remains unclear.
\item  In comparison to experimental data for $x=0.25$ it
is evident, that the width of the diffuse scattering is too small and the fourfold star like structure is missing~\cite{gaw13-165121}. However, more recent experimental data are in better agreement~\footnote{Very recent experimental data on a high quality x=0.25 sample by A. Komarek, private communication, clearly shows a very narrow feature, which is in better agreement with the calculation result shown fig.~\ref{qq-x0p25}}. 
\item For $x=0.4$ there is too strong magnetic diffuse scattering around (0.5 0.5 0) in comparison to the experiment~\cite{cwik09-057201,drees13-3449}.
\end{itemize}.

In the following sections we report an attempt to improve the understanding of anisotropy and of the diffuse scattering 
within the framework of a frustrated checkerboard charge order model.

\begin{figure}[htbp] 
  \centering
     \includegraphics[width=0.5\textwidth]{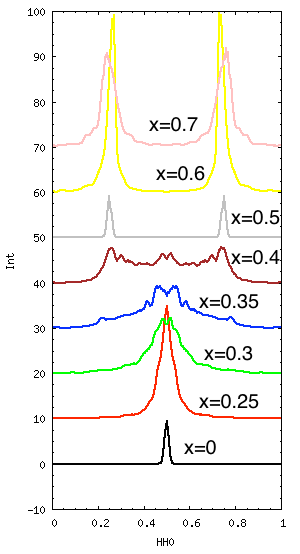}
  \caption{Simulated elastic magnetic neutron pattern along hh0 for different $x$. }
  \label{r02m1-hh0}
\end{figure}

\section{Investigation of the role of single ion anisotropy in the formation of the excitation gap}

The role of single ion anisotropy in \lsco was investigated. In \lsco it is known that
the magnetic moments order perpendicular to the $c$-axis. This anisotropy can be produced by 
a crystal field $B_2^0$ parameter as described in section~\ref{magprop}. Yet, such anisotropy will
not lead to a gap in the excitation spectra (13~meV for $x=0$, 3~meV for $x=0.5$), because
the moment direction within the $ab$ plane is not fixed leading to a Goldstone mode.

One possible cause of additional single ion anisotropy is the low temperature orthorhombic (LTO)
 distortion of \lsco~\cite{cwik09-057201} with a unit cell approximately
$\sqrt{2}\times \sqrt{2}$ the tetragonal unit cell. This orthorhombic symmetry may lead (among others) to an additional
crystal field parameter $B_2^2(s)$. In order to reproduce the experimental gaps, we  set this
parameter $B_2^2(s)=-3.97$~meV  for $x=0$ and $B_2^2(s)=-1.63$~meV for $x=0.5$. The
calculated spectra for x=0 and x=0.5 are nearly identical with those shown in figs.~\ref{qq-x0p00} and
\ref{qq-x0p50}.

Fig.~\ref{r02m1-hh0} shows the results of magnetic elastic scattering using an average of 4 charge order grids, demonstrating
that there is no improvement in the description of the experimental data with respect to the previous model.

\begin{figure}[htbp] 
  \centering
     \includegraphics[width=0.4\textwidth]{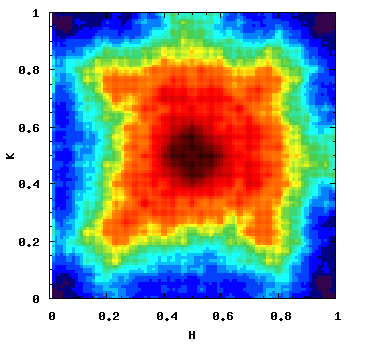}
  \caption{Simulated elastic magnetic neutron pattern in the (hk0) plane for x=0.3 including local variations in Co-O distance. }
  \label{r03m2-hk0}
\end{figure}

Moreover, despite it's success in describing the gap the real cause of anisotropy has to be more complex than
a simple introduction of an orthorhombic $B_2^2(s)$ parameter: (i) for $x=0.5$ no orthorhombic distortion has
been observed at any temperature (ii) also the parent phase $x=0$ is not orthorhombic at low temperature, it
 exhibits a low temperature tetragonal phase (LTT). Characteristic
 is the tilt of the octahedron in the low temperature
tetragonal (LTT) and low temperature orthorhombic (LTO) phases. The tilts about the [1,~-1,~0] direction form 
an antiferrodistortive structural arrangement~\cite{yamada89-2336}. These tilts produce more complex single ion
anisotropy than just a nonzero $B_2^2(s)$ parameter.  Therefore, in the next step single ion anisotropy is described by a tilt
of the octahedron. The additional nonzero crystal field parameters were estimated from the point charge
model and doubled to effectively account for bonding effects. The tilt was assumed to vary depending on
the local concentration of \cppp ions. For $x=0$ a tilt of 12 deg and for $x=0.5$ a tilt of 10 deg is found to
yield the experimental anisotropy gaps.  

\section{Investigation of the role of local variations in Oxygen Positions}

It is a widely accepted assumption that for \cppp based oxides with octahedral coordination the cobalt-oxygen (Co-O) distance
will determine the spin state of \cppp. The prototype system LaCoO$_3$ with a temperature induced low spin (LS) to high spin (HS) transition
exhibits a Co-O distance of 1.925~\AA~\cite{radaelli02-094408}. The Co-O distances in \lco are above this critical value and thus doping
with Sr  should induce a HS state of \cppp in contrast to the widely used assumption, that \cppp in \lsco is in it's LS nonmagnetic state.
%
Another peculiarity of \lsco is that fact, that on doping with Sr the lattice shrinks, although Sr$^{2+}$ has a larger ionic radius than 
La$^{3+}$. 

From these observations it seems possible, that there are local variations in Co-O distances depending on the ionic radius of the 
Co ion. Actually such variations are needed in order to justify the steric repulsion of adjacent \cpp used in
the model for the checkerboard charge order (see section~\ref{comodel}). According to crystal field theory a local variation in Co-O distance
should also induce a corresponding crystal field anisotropy. Moreover, local variations of oxygen positions might also affect the
hopping and thus influence the exchange interactions between the magnetic ions. In a series of model calculations these
effects were investigated and the footprints in the diffuse magnetic scattering were identified. An example of such a 
simulation is shown in fig.~\ref{r03m2-hk0}. 

\subsection{Local variations in Crystal Field Anisotropy}

The local additional crystal field anisotropy of \cpp ions induced by an asymmetric \cppp neighbourhood
is described by a local variation in the crystal field parameter $B_2^2$, which was estimated from the point charge model assuming a shift
of oxygen positions of 0.13~\AA\, according to the difference in ionic radii of \cpp (0.88~\AA) and \cppp  (0.75~\AA\, for the HS state, 0.69~\AA\, for LS) . If there is only one \cppp neighbour  this anisotropy will tend to align moments perpendicular to the direction of the \cppp neighbour. In this way the nano clusters of the parent ($x=0$) compound will tend to have their moments aligned
parallel to their major  axis. For example if such a cluster is elongated along (100), moments will also tend to point along (100) and the $J(100)$ interactions will induce a propagation of (0.5 k 0), k depending on the actual shape of the cluster. Due to the polarisation factor the intensity will be larger at (0.5 1 0) than at
(0.5 0 0) leading to an asymmetry in the diffuse scattering shown in fig.~\ref{r03m2-hk0} - the ''star'' is not symmetric, but there is more intensity  around (0.5 0.8 0) and (0.8 0.5 0) than around (0.5 0.2 0) and (0.2 0.5 0), respectively. However, such an asymmetry in the diffuse scattering has not been found in the experiment~\cite{gaw13-165121,drees13-3449}. Thus local variations of oxygen positions of the order of 0.1~\AA\, are unlikely in \lsco.

\subsection{Local variations in Exchange Interactions}

 Moreover, some effect of different Co ion size on exchange interactions between a nearest neighbour
pair of \cpp ions was taken into account in the simulation shown in fig.~\ref{r03m2-hk0}: if on one side of the pair there is a \cppp nearest neighbour, the interaction  was reduced from -9.5 to -2.5 meV, if on both sides there are \cppp nearest neighbours it is taken to be +0.5 meV. Thus bonds perpendicular to the major axis of a parent phase nano cluster are weaker and the cluster can be viewed approximately as a series of one dimensional chains  with nearest neighbour interaction $J(100)=-9.5$~meV along the major axis. This will lead to diffuse intensity stripes along (0.5 k 0) and (h 0.5 0), which are also not observed in the experiment. Note that in the simulation shown in  fig.~\ref{r03m2-hk0} these intensity stripes are visible only at (0.5 1 0) and (1 0.5 0) and not at (0.5 0 0) and (0 0.5 0) because of the local variations of single ion anisotropy described above.
Thus also the absence of local variations in exchange interactions contradict the assumption of local variations of oxygen positions in \lsco.

\begin{figure}[htbp] 
  \centering
     \includegraphics[width=0.41\textwidth]{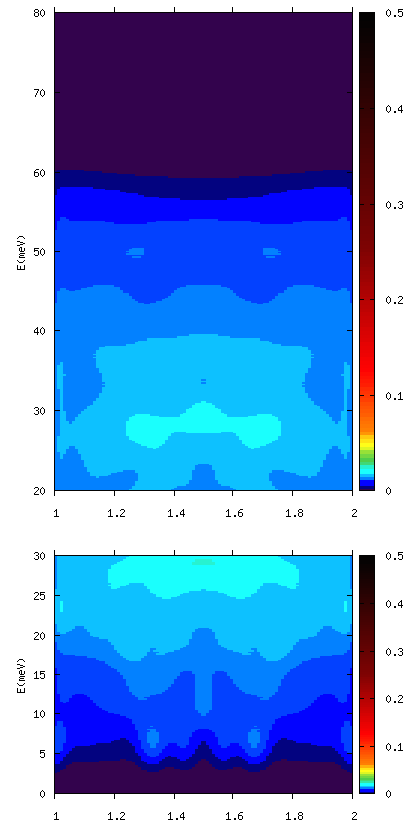}
     \includegraphics[width=0.41\textwidth]{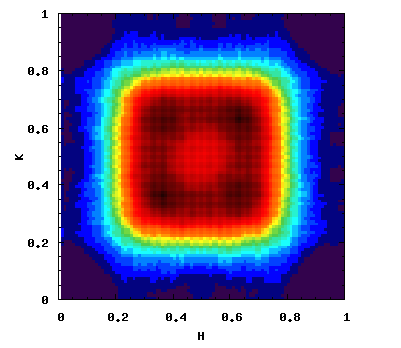}
  \caption{Simulated inelastic (top: along (h 0.5 0), middle: along (hh0)  and elastic (bottom: (hk0) plane)  magnetic neutron pattern for x=0.25 }
  \label{r04mJ-x0p25}
\end{figure}

\section{Spin nematic interactions}

Up to now we have assumed a nonzero crystal field anisotropy in the $ab$ plane for all $x$ in \lsco.
However, because of the absence of a tilt of the octahedron this assumption is not justified for $x=0.5$.
Thus it is necessary to look out for other possible sources of anisotropy, which could lead to the observed
anisotropy gap. Recently, Soda et al.~\cite{soda14-127205} pointed out that spin nematic interactions of the form 
$ -1/2 \sum_{ij} K(ij) O_2^2(s)(\hat {\mbf L}_i) \cdot  O_2^2(s)(\hat {\mbf L}_j) $ may induce an anisotropy gap. In the following
we used such iterations with $K(200)=0.05$~meV  and found that in this way the anisotropy gap
for $x=0.5$ can be successfully described. Thus in our final approach to anisotropy investigation  spin nematic interactions are used. $K(200)$ is put
nonzero only, if a
  \cppp ion is at the intermediate (100) position mediating the interaction.

In addition, depending on the number of adjacent \cppp ions the tilt angle induced anisotropy is 
linearly scaled from 12 deg (zero adjacent \cppp ) to 0 deg (4 adjacent \cppp ions). 

\section{Increasing Frustration by Variations in (110) and  (200) interactions}

The discrepancy in diffuse scattering between experimental data and results of the model calculation
are (i) the width and anisotropy of the diffuse scattering for $x=0.25$ and (ii) the intensity around (1/2 1/2 0)
for $x=0.4$. The description of these two features of the experimental data may be significantly improved
by an increase in frustration via introducing  variations in the exchange couplings $J(110)$ and $J(200)$. 

In the studies of 
Babkevich~\cite{babkevich10-184425} and Helme~\cite{helme09-134414} it was found
that  $J(110)$ is zero for $x=0$ and $x=0.5$ - yet it is unclear, why the simple estimate
based on exchange paths involving a single \cppp bonding orbital $J(110)=2 \cdot J(200)$ does not apply.
Here we introduce nonzero $J(110)=-2$~meV only, if exactly one of the two $(100)$ and $(010)$ adjacent neighbours is
\cppp and the other is \cpp. This rule will yield zero $J(110)$ for the parent compound $x=0$ and
also for the checkerboard charge order $x=0.5$. The size of $J(200)$ is assumed to vary proportional
to the amount of \cpp ions in the neighbourhood, i.e. we choose  $J(200)=-1.4$~meV if 0,1 or 2 out of 6 NN are \cpp,
$J(200)=-3.5$~meV if 3 or 4 out of 6 NN are \cpp and $J(200)=-7.0$~meV if 5 or 6 out of 6 NN are \cpp.

Fig.~\ref{r04mJ-x0p25} shows the result of such a simulation. The elastic diffuse magnetic scattering shows
a fourfold pattern, which does not resemble  the perfect ''star'' observed in the experiment~\cite{gaw13-165121}.
The inelastic pattern for $x=0.25$ has no similarity to the hourglass, because the huge magnetic frustration leads to a lot of broad diffuse scattering
and prevents a resonance like peak.

\begin{figure}[htbp] 
  \centering
     \includegraphics[width=0.4\textwidth]{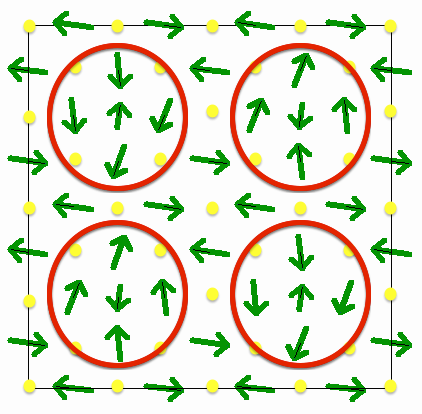}
     \includegraphics[width=0.4\textwidth]{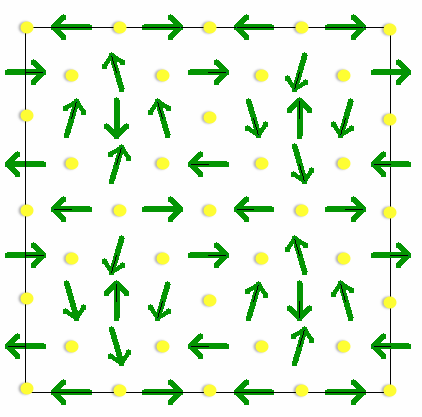}
  \caption{Regular doped checkerboard charge order pattern corresponding to x=0.4375. Red circles indicate the magnetic clusters
which consist of 5 strongly interacting \cpp ions. The calculated spin structure using a cluster approach (top) is compared to that
obtained with a standard single ion calculation (bottom).  }
  \label{r05c1-x0p4375CO}
\end{figure}

\begin{figure}[htbp] 
  \centering
     \includegraphics[width=0.4\textwidth]{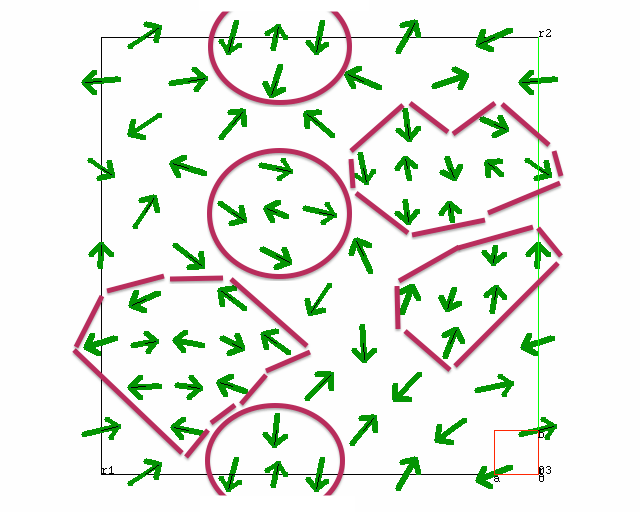}
  \caption{Magnetic order on a 10x10 frustrated checkerboard charge order pattern corresponding to x=0.42. Red circles indicate the magnetic nano clusters.
 The calculated spin structure using a cluster approach is shown.}
  \label{r05c3-x0p42_s10_CO}
\end{figure}

\section{Beyond Single Ion theory - some Cluster Calculations}

In order to investigate the role of quantum fluctuations it is possible to make a first step by forming out of strongly interacting adjacent \cpp ions a 
magnetic cluster. For computational reasons we revert to the simplest model described by~\cite[Drees et al.]{drees14-6731}, which has been shown to predict an hourglass spectrum  (anisotropic effective S=1/2 without
 in plane anisotropy $\langle S_x \rangle = \langle S_y \rangle = 1.5, \langle S_z \rangle = 1.3 $, $J(100)=-5.8$~meV, $J(200)=-0.85$~meV).
If checkerboard charge order would not freeze out some stochastic charge configuration it might be possible to observe regular charge order pattern, for example
in a $8 \times 8$ superstructure as shown in fig.~\ref{r05c1-x0p4375CO} and corresponding to $x=7/16=0.4375$. For this pattern the magnetic properties were calculated (i) using exact diagonalisation of the five-ion clusters indicated by the red circles  in fig.~\ref{r05c1-x0p4375CO}(top) and doing a mean field random phase approximation (MF-RPA)
for the weak interactions between the clusters and the remaining \cpp ions~\cite{jensen09-014406}, the calculated magnetic moments are indicated by the arrows in fig.~\ref{r05c1-x0p4375CO}(top). For comparison,
(ii) all \cpp ions were treated separately, see fig.~\ref{r05c1-x0p4375CO} (bottom).  The cluster-approach (i) yields
a smaller magnetic moment on the central \cpp ion of the cluster.

\begin{figure}[htbp] 
  \centering
     \includegraphics[width=0.45\textwidth]{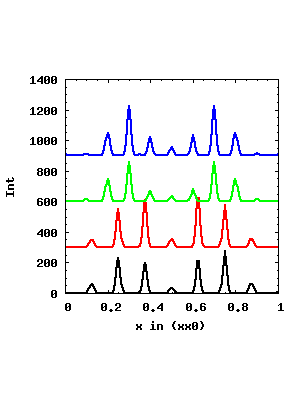}
\caption{Diffraction pattern along (xx0) obtained by averaging 8 grids. A calculation with magnetic nano clusters  consisting of up to 11 ions is
compared to a calculation using single \cpp ions on the same frustrated checkerboard charge order grids
corresponding to $x\sim 0.42$.  From top to bottom: (i) grid dimension 10x10, no nano clusters; (ii) same grids  as (i) but with nano clusters; (iii) grid dimension 8x8, no nano clusters; (iv) same grids as (iii) but with nanoclusters.}
  \label{r05c3-hh0}
\end{figure}

\begin{figure}[htbp] 
  \centering
     \includegraphics[width=0.4\textwidth]{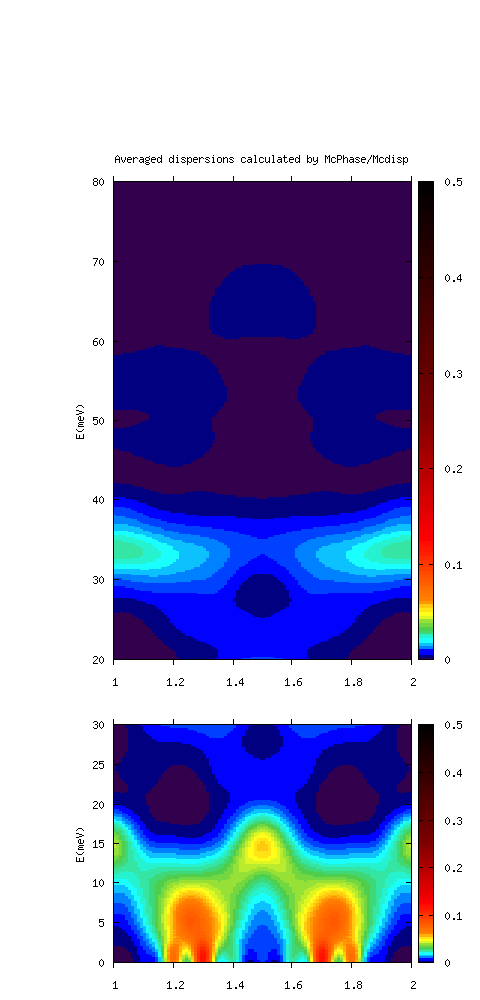}
  \caption{Averaged spectrum (top) of 8 frustrated checkerboard charge order grids  (dimension 10x10) with nano clusters as used in fig.~\ref{r05c3-hh0} (top). top: along (h 0.5 0), bottom: along (hh0)   }
  \label{r05c3-x0p42_s10}
\end{figure}

\begin{figure}[htbp] 
  \centering
     \includegraphics[width=0.4\textwidth]{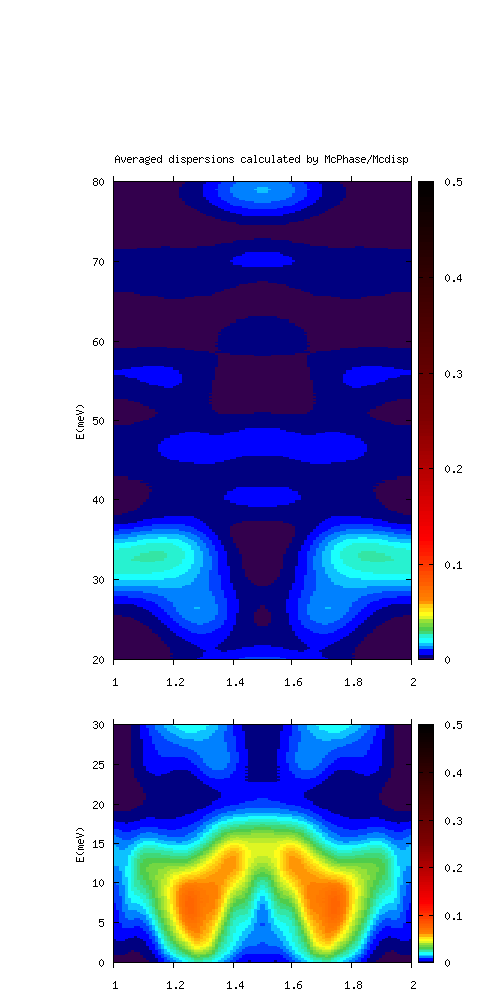}
  \caption{Averaged spectrum of 8 frustrated checkerboard charge order grids (dimension 10x10, as used  in fig.~\ref{r05c3-hh0} top), however calculated using the standard 
MF-RPA approach without formation of magnetic nano clusters . top: along (h 0.5 0), bottom: along (hh0)   }
  \label{r05c3-x0p42_s10_noclust}
\end{figure}

Therefore it is likely, that quantum fluctuations will tend to suppress (1/2 1/2 0) correlations for $x \sim 0.4$ and that the high intensity, which has been
calculated around (1/2 1/2 0) is an artefact of the single ion spin wave or MF-RPA approach. In order to see this effect more quantitatively the model was extended to a grid size of $10 \times 10$ and clusters consisting of up to 11 ions.  An example of such a grid with clusters is shown in fig.~\ref{r05c3-x0p42_s10_CO}. Also here the
tendency of smaller moments within the nanocluster can be seen.
Fig.~\ref{r05c3-hh0} compares a diffraction pattern along (xx0) obtained by averaging 8 grids with magnetic nano clusters (up to 11 ions) to  a calculation using single \cpp ions on the same frustrated checkerboard charge order grids
corresponding to $x\sim 0.42$. The reduction of the intensity around the center of the plot is clearly visible and stems from nano cluster quantum effects, which are not accounted for in the mean field approach using single \cpp ions.

It has still to be investigated, if also the formation of the
resonance like peak is suppressed by such quantum fluctuations and thus the explanation of the hourglass spectrum given in~\cite{drees14-6731} has to be 
revised. Therefore the spectral response was calculated for the same grids, which were used to generate fig.~\ref{r05c3-hh0} and is shown in
fig.~\ref{r05c3-x0p42_s10} and fig.~\ref{r05c3-x0p42_s10_noclust}. The tendency to form a resonance and an hourglass  spectrum can
clearly be seen in the calculation using nano clusters. 

Unfortunately extending this type of calculation to lower concentrations $x<0.4$ soon runs into the limits of available computational power, because the
nano cluster size gets too large. Nevertheless we expect, that the tendency to reduce the central moments of the nano clusters will persist and will
lead to a broadening of the magnetic response down to possibly $x=0.25$. This might explain, why at such small $x$  a hourglass spectrum has been observed
in the experiment, however,  without any clear footprint of stripe charge order.

\section{Conclusion}

Numerical calculations of the diffuse and inelastic neutron scattering in comparison with experimental data on \lsco identify
spin nematic interactions as a source of anisotropy for $x=0.5$.
Magnetic quantum nano cluster based calculations of spectra on a series of frustrated checkerboard charge order grids for $x\sim0.42$ show, that quantum fluctuations will  destroy short range magnetic order of \cpp rich clusters thus reducing neutron intensity around (1/2 1/2 0) and possibly leading to an hourglass type of response without charge order stripes for $x$ down to 0.25. 

Summa summarum there remain two puzzles in \lsco: (i)  the \cppp ions seems to be in a LS state, although the Co-O distance is above the threshold for
a HS-LS transition. (ii) Looking at experimental diffuse magnetic scattering data no features for local variations of the in plane oxygen positions of the order of 0.1~\AA\, due to doping can be identified. Therefore the suitability of a nearest neighbour steric repulsion of adjacent \cpp ions  in the charge order model may be questioned.

\begin{acknowledgements}
The fruitful scientific discussions with Hideki Yoshizawa during my stay at ISSP are highly appreciated. The help of Naoki Kawashima in bringing
McPhase to the supercomputer is greatly acknowledged. I wish to thank  Jens Jensen for his  help in testing  the cluster module of McPhase and Alexander C. Komarek for sharing his recent experimental data on a sample with $x=0.25$.
The computation in this work has been done using the facilities of the Supercomputer Center, the Institute for Solid State Physics, the University of Tokyo.
\end{acknowledgements}

\bibliographystyle{physrev}
\bibliography{li160111}

\end{document}